# Particle dynamics and pattern formation in a rotating suspension of positively buoyant particles

Sudarshan Konidena, Jonghoon Lee[+], K. Anki Reddy[*] and Anugrah Singh[*]

Department of Chemical Engineering, Indian Institute of Technology, Guwahati, India.
+ Materials and Manufacturing Directorate, Air Force Research Laboratory, Wright-Patterson Air Force Base, Ohio, USA.

Numerical simulations of positively-buoyant suspension in a horizontally rotating cylinder were performed to study the formation of radial and axial patterns. The order parameter for low-frequency segregated phase and dispersed phase is similar to that predicted for the settling suspension by J. Lee, and A. J. C. Ladd [*J. Fluid Mech.,* 577, 2007], which is the average angular velocity of the particles. The particle density profiles for axial bands in the buoyancy dominated phase shows an amplitude equivalent to the diameter of the cylinder. Axial density profiles show sinusoidal behaviour for drag dominant phase and oscillating sinusoidal behaviour for centrifugal force dominant phase. Results also indicate that the traveling bands are formed as a consequence of the inhomogeneous distribution of particles arising from a certain imbalance of drag, buoyancy and centrifugal forces. In the centrifugal limit, particles move towards the centre of the cylinder aggregating to form a dense core of particles with its axis coinciding with that of the rotating cylinder, a behaviour which is in contrast to the sedimenting particles. The particle distribution patterns obtained from the simulations are found to be in good agreement with the experiments of Kalyankar *et al.* [*Phys. Fluids*, 20, 2008].

* Corresponding Authors: anki.reddy@iitg.ernet.in, anugrah@iitg.ernet.in

## Nomenclature

| | |
|---|---|
| $a$ | radius of particle |
| $d$ | mean interparticle separation |
| $g$ | acceleration due to gravity |
| $\mathcal{G}$ | Green's function for the Stokes equation |
| L | length of the cylinder |
| $m$ | mass of the particle |
| $m_B$ | buoyancy corrected mass of the particle |
| $n$ | number of particles |
| $n_0$ | average particle concentration |
| $n_p$ | number of particles at steady state |
| $N_0$ | number of particles at t = 0 |
| R | radius of the cylinder |
| $Re_f$ | flow based Reynolds number |
| $Re_p$ | particle based Reynolds number |
| $u_0$ | settling velocity of the particle |
| $u_f$ | floating velocity of the particle |
| $u_r$, | radial velocity of the particle |
| $u_\theta$ | angular velocity of the particle |
| $s_i$ | particle surface |
| V | volume of spherical particle |
| μ | viscosity of the fluid |
| ν | kinematic viscosity of the fluid |
| Ω | rotational velocity of the cylinder |
| $\rho_f$ | density of the fluid |
| $\rho_p$ | density of particle |

## I. INTRODUCTION

Pattern formations like periodically spaced ripples on sand or water, patterns on basin floor, Fibonacci patterns on leaves etc. have been of interest to mathematicians, physicists and researchers of various disciplines due to their omnipresence in nature. One can encounter an exhibition of such rich patterns in the flow of both dry and wet granular matter under a different set of conditions. Study of the dynamics of suspensions consisting of micrometre and sub-micrometre particles holds many applications in industrial, domestic and environmental avenues. Although being ubiquitous, the flow of non-colloidal suspensions itself is not completely understood. This dearth of knowledge on suspension flows can be ascribed to the mathematical complexity associated even with dilute systems.

Flows in a horizontal rotating-cylinder or a Taylor-Couette device also display various non-equilibrium patterns. Experiments with dry granular particles in rotating cylinders revealed that the particles separate into a series of bands and get stacked along the horizontal axis[1], these experiments probed the evolution of axial bands and its dependence on the rotational velocity of the horizontal cylinder. Other experiments showed that dry granular matter segregates by mass and size in the cylindrical geometry[2]. In dry granular matter the dominant interactions are inelastic frictional collisions between the particles, nevertheless, wet granular slurries too exhibit particle segregation in which the particle interactions are mediated by the interstitial fluid. For a range of rotational frequencies, suspensions exhibit the phenomenon of axial segregation which appears like stripes of particles interlaced in pure fluid[3]. The phenomenon of non-equilibrium patterns along the radial and axial directions for particle laden flows in horizontal rotating cylinders has been of interest to researchers owing to diversified applications. These patterns in horizontal cylinders appear in two cases, firstly, with the interaction of a rotating bulk suspension of particles with a free surface and secondly, in the case of fully filled cylinders suspending non-neutrally buoyant particles. Moreover, these two types can be further subdivided with reference to the nature of the suspended particles. Experimental works revealed that for fully-filled cylinders to exhibit axial segregation, the suspended particles should necessarily be non-neutrally buoyant. Though, a fair amount of investigation has been done on both the experimental and numerical fronts, a unifying mechanism to explain axial segregation is yet to be arrived at. Significant contributions from various analyses (both experimental and theoretical) are briefed below to arrive at the objective of the current work.

Bands of particles are reported to be observed in partially filled Taylor-Couette devices with the inner cylinder rotating and concentration of particles being as low as 1% to as high as 65%. The banding patterns usually take considerable time to appear for less concentrated suspensions and appear soon if the concentration of particles in the suspension is maintained high. In another independent study experiments were conducted in a Taylor-Couette system for different fill-fractions and neutrally buoyant suspended particles.[3,4] In addition to reporting the occurrence of bands due to axial segregation of particles this work also reiterated the proposition of Boote *et al.*[5] that axial banding does not result from the presence of the wave front as suggested by Melo[6]. Later Joseph *et al.*[7] performed experiments in particle-laden rimming flow with floating particles. Particles were either less dense than the carrier fluid or hydrophobic allowing them to float on the liquid-air interface.

A theoretical explanation to the formation of granular bands with the help of mathematical model was first attempted by Govindarajan *et al.*[8]. Based on the concept on shear induced migration of particles and a concentration dependent viscosity in the presence of a free surface, the approach claimed to have provided a qualitative explanation to the phenomenon observed in Refs. 3 and 4. However, the work done by Timberlake and Morris[9] revealed that the time scale for shear-induced migration to affect the particle clustering dynamics is nearly 40 times lesser than the observed experimental time scale. Later, Jin and Acrivos[10,11] examined the stability of the suspension to axial disturbances in the concentrations of the particles by performing a linear stability analysis. In their work it was assumed that the radial patterns of the suspension precede axial segregation. The investigations done by Jin and Acrivos[11] concluded that the underlying cause for the instabilities in the particle distribution to the axial perturbations could be surface tension. Therefore the selection of wavelength of the most rapidly amplified disturbance could be attributed to surface tension. Duong *et al.*[12] used the variable viscosity approach to report the existence of alternating solid-like and liquid-like states in highly concentrated suspensions. Raiskinmaki *et al.*[13] however coupled the application of direct numerical simulations with variable viscosity approximation to comprehend formation of clusters in Couette flows. Theoretical works till date could not completely establish a comprehensive mechanism behind suspended particles leaving the pool of liquid to get radially segregated inducing particle clustering along the axis of the cylinder.

Apart from probing the formation of axial bands in partially-filled cylinders a few experiments were performed with dilute suspensions entirely filling the rotating cylinder. One of the major differences between the cases of suspension partially and fully filling the cylinders is the existence of a free surface. The segregation mechanisms for the partially filled suspensions are governed strongly by the free-surface dynamics[10]. However, even with the fully-filled rotating cylinders apparently similar band formation was observed[14]. To start with, it is to be noted that most of the studies which explored the parameter space of axial banding phenomenon for the case of fully filled cylinders were conducted with sedimenting particles. Lipson[15] reported alternating concentration bands of particles along the axial direction of the rotating cylinder. His experiments were directed at measuring the average spacing of bands as a function of the ratio of length and radius of the cylinder. Millimetre sized particles which constitute the suspension are responsible for the high particle based Reynolds numbers ($Re_p = 2a\,u_0/\nu$ where $a$ is the particle radius, $u_0$ is the settling velocity of an isolated particle and $\nu$ is the kinematic viscosity of the carrying fluid) ranging from 6.5-735 in their experiments whereas the effect of high rotational frequencies is reflected in the high flow based Reynolds number ($Re_f = \Omega R^2/\nu$, here $\Omega$ is the rotational velocity of the cylinder and R is radius of the cylinder).

Breu *et al.*[16,17] performed extensive studies with low viscosity interstitial fluids suspending glass beads. These studies were performed at high rotational frequencies of the cylinder. They observed instabilities which highlight the hysteric character of the transition states as the rotational velocity is lowered. The axial bands also periodically expand and shrink along the axis depending on the rotation rate of the cylindrical drum. Particle based Reynolds number for the experimental conditions of Breu *et al.*[16,17] is $Re_p \sim 20$ clearly implicating that inertia plays a significant role for band formation.

Matson and co-workers[18-20] used carrier fluids with viscosities ranging from 0.05-1 $cm^2/s$. A series of experiments were performed to identify the phase space of the various non-equilibrium states. For their experimental conditions, in the low rotational frequency limit and for various fluid viscosities, the particle based Reynolds number $Re_p < 0.1$. Therefore, in this limit inertia is negligible and without loss of generality one can apply the Stokes flow approximation. At low rotational frequencies, the alternate concentrated particle stripes were sinusoidal in nature while at higher frequencies they were more pronounced but asymmetric.

Lee and Ladd[21] proposed a theory to comprehend the underlying particle dynamics for the non-equilibrium patterns observed by Matson *et al.*[18]. They claimed that for a suspension of centrifuging particles, different centrifugal forces on particles at different radial positions produce an attractive interaction and relative motion between the particles along the axial direction. The relative motion between the particles in turn amplifies the axial density fluctuations forming concentrated particle bands along the rotational axis. Their claim also suggests that for a suspension of non-Brownian buoyant particles, differential centrifugation produces a repulsive interaction between the particles which stabilizes any axial density fluctuations if present. To be noted is that Lee and Ladd[21] approximated the cancelling field of cylindrical wall with a flat wall. However, it was later realized that the correct treatment of the no-slip cylindrical boundary condition should exactly nullify any interaction between ring stokeslets due to differential centrifugation, which, therefore, cannot amplify any axial density perturbation[22,23]. In their follow-up works[22,24] Stokeslet simulations were used to reproduce the low frequency patterns obtained in the experiments done by Matson *et al.*[18] and reported that the axial banding might be a result of the secondary flow caused by sedimenting clusters during the dynamic phase transition in the radial plane.

Motivated by the theory put forward by Lee and Ladd[21], Kalyankar *et al.*[25] conducted experiments with buoyant non-Brownian suspension and performed a comparative study with the settling suspension. Except for the densities of the suspended particles, most other parameters like the cylinder diameter, length, particle sizes, etc., were similar to those used in Matson *et al.*[18-20]. It was found from their experiments that when the action of gravity overcomes centrifugal forces, non-equilibrium states for the buoyant and the settling suspensions can be correlated. Kalyankar *et* *al.*[25] used glycerine-water mixtures as the carrier fluid with kinematic viscosities ranging from 0.25-1 cm$^2$/s. The particle based Reynolds number Re$_p$ ~ 0.01-0.1 while the flow based Reynolds number at the lowest and highest frequencies of rotation Re$_f$ ~ 1-100. However, for most of the axial band phases exhibited in their experiment Re$_f$ ~ 1. As many as nine independent non-equilibrium patterns were observed for the buoyant particle system. Among those, a low frequency axial segregation was also reported, in contradiction to the suggestion of Lee and Ladd[21] based on underestimated cancelling field.

The present work is aimed at studying the dynamics of radial and axial patterns in rotating suspension of positively buoyant particles. As the experimental conditions of Kalyankar *et al.*[25] are similar to those of Matson *et al.*[18-20], we followed the simulation method detailed in Lee and Ladd[24] to investigate the low frequency banding phases that were reported. In this method, the hydrodynamic flow fields around particles are approximated by Stokeslets and summed up with the flow field generated by imposing a no-slip boundary condition on the cylinder surface. The simulations reproduced the transference of axial bands of high and low concentration, which occurs at a relatively high frequency compared to those reported by Lee and Ladd[22].

## II. SIMULATION METHOD

In this section, the simulation methodology adopted in the current work is described briefly. We consider a dilute suspension of monodisperse solid spheres of radius '*a*' in a horizontal cylinder of radius *R*, rotating at an angular velocity Ω. The equation of motion of a particle *i* with mass *m*, is given by,

$$m \frac{d\mathbf{u}_i}{dt} + 2m\mathbf{\Omega} \times \mathbf{u}_i = m_B \mathbf{g} - m_B \mathbf{\Omega} \times (\mathbf{\Omega} \times \mathbf{r}_i) + \iint \boldsymbol{\sigma} d\mathbf{s}_i \qquad (1.1)$$

where, $\mathbf{u}_i$ is the velocity of particle *i*, $m_B$ is the buoyancy corrected mass, $\boldsymbol{\sigma}$ is the fluid stress integrated over the particle surface $\mathbf{s}_i$. The last term on RHS of the above equation is contribution of the drag force that depends on positions of the surrounding particles.

The suspending fluid in the cylinder is assumed to be Newtonian and the inertial forces are neglected based on the experimental conditions of Kalyankar *et al.*[25]. On applying this approximation to the eq. (1.1), it reduces to

$$\mathbf{F}_j = m_B \mathbf{g} + m_B \Omega^2 \mathbf{r}_j = \xi [\mathbf{u}_j - \mathbf{u}(\mathbf{r}_j)] \qquad (1.2)$$

where, $\mathbf{u}(\mathbf{r}_j)$ is the fluid velocity at particle location $\mathbf{r}_j$, $m_B = (\rho_f - \rho_p)Vg$; *V* is the volume of the sphere and $\xi = 6\pi\mu a$.

A) Single Particle Dynamics

Before proceeding into understanding the collective behaviour of the suspended particles in the viscous fluid an attempt is made to understand the dynamics of a single particle. This study is performed to classify the system based on the strength of gravitational, centrifugal and drag forces acting on it. A positively buoyant particle placed in a viscous fluid which rotates along its horizontal axis experiences a buoyancy corrected gravitational force and a centrifugal force given by $-m_B g \hat{y}$ and $m_B \Omega^2 r \hat{r}$ respectively ($\hat{y}$ and $\hat{r}$ are the unit vectors in y and r directions respectively). In isolation, the resultant of these forces gives rise to a rising velocity $u_f \hat{y}$ and a centrifuging velocity $u_c \hat{r}$ where $u_f = m_B g \xi^{-1}$ and $u_c = m_B \Omega^2 r \xi^{-1}$. On solving the instantaneous force balance equation the velocity of the isolated particle when resolved into the cylindrical coordinate system is given by

$$u_r = u_f \left(\frac{r}{D_1 R} + \sin\theta\right), \qquad u_\theta = u_f \left(\frac{r}{D_2 R} + \cos\theta\right) \tag{1.3}$$

where, $D_1 = g/\Omega^2 R$ and $D_2 = u_f / \Omega R$ are dimensionless numbers which describe the relative magnitudes of the gravitational, centrifugal and drag forces acting on the isolated particle as defined by Lee and Ladd[24]. Trajectories on which the radial and the angular velocities become zero for the particle under consideration can be obtained by equating $u_r$ and $u_\theta$ to zero. The loci so obtained are circles $C_1$ and $C_2$ with diameters $D_1 R$ and $D_2 R$ as shown in Fig. 1(a) which intersect at point P (unstable equilibrium). At very low rotational velocities $D_2 > 1$ hence the particle reaches the point of stagnation A in Fig. 1(b). Locations of P and A differ from Lee and Ladd[24] owing to the reversal of the direction of centrifugal force.

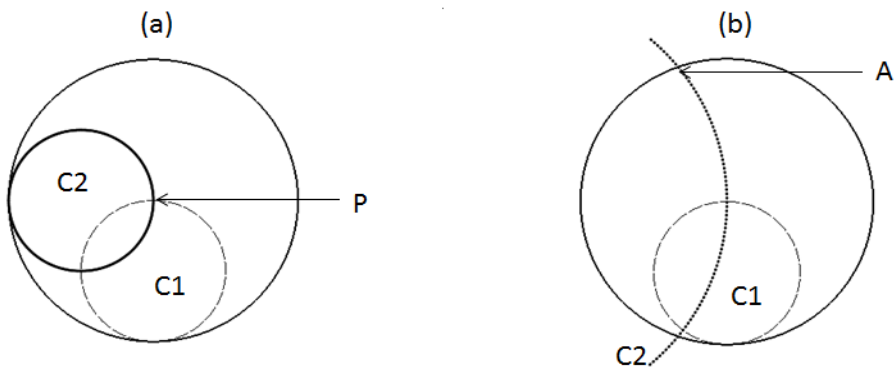

FIG. 1: The circles $C_1$ and $C_2$ describe the locus of zero radial and angular velocities respectively. (a) corresponds to the drag dominant phase with the point P indicating the dynamical centre of the system and (b) corresponds to centrifugal dominant phase, here the point A indicates the stagnation point.

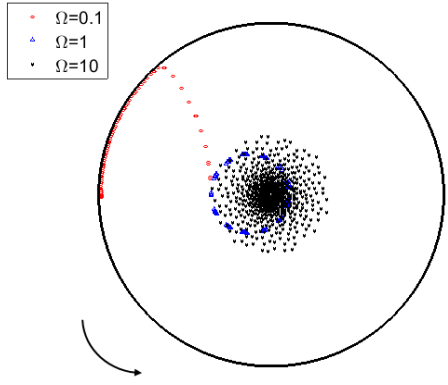

FIG 2: Particle trajectories at different velocities of the cylinder rotating counter-clockwise as indicated by the arrow. Initial position of the particle in all three cases is $(r, \theta, z) = (0.5, 0, 0)$. $\Omega = 0.1$: drag dominant regime, the particle rises to reach the cylinder wall and gets dragged down because of rotation. $\Omega = 1$: balance of gravity, drag and centrifugal forces imposing a closed trajectory to the particle. $\Omega = 10$: centrifugal force dominant regime, the particle swirls inward to the centre.

The trajectories of a single particle for three different rotational frequencies shown in Fig. 2 signify the balance of forces acting on the particle. As discussed earlier, the particle rises up and moves along the wall for very low rotational frequency (see the Fig. 2a) and swirls towards the centre due to high centrifugal force as shown in Fig. 2(c). Nevertheless, it has a closed trajectory in Fig. 2(b) implicating a balance of forces acting on the system. In lieu of the above arguments it is justified to classify the system into three regimes namely; a) buoyancy dominated b) a balance between buoyancy and centrifugal forces and c) centrifugal force dominating regime.

B) Stokeslets confined in a cylinder

In the case of multi-particle dynamics, the hydrodynamic interactions also play a role in addition to the forces detailed in single particle dynamics. Since the interest lies in comprehending the collective behaviour of particles in the suspension, the additional contribution to the particle velocities from interactions is incorporated. The approach followed to understand collective particle dynamics is similar to the molecular dynamic simulation. First, the particle positions are initialised using Monte-Carlo method to generate random initial configuration for the particles. The initial configuration is generated by maintaining the volume fraction of the particles $\phi \sim 0.02$ pertaining to the experimental conditions. In the second step the velocity field for the system is computed from the equation (2.2); this is done by approximating the particles as Stokeslets and summing up with the flow field due to the cylinder surface. Formulation of this velocity field is stated briefly below.

Starting from the fundamental solution for Stokes flow termed as the Stokeslet, we can construct a general solution for the dynamics of point particles confined in a cylindrical boundary. A schematic representation of the cylinder with its rotational axis aligned along the z-axis, and fully filled with suspension is shown in Fig. 3.

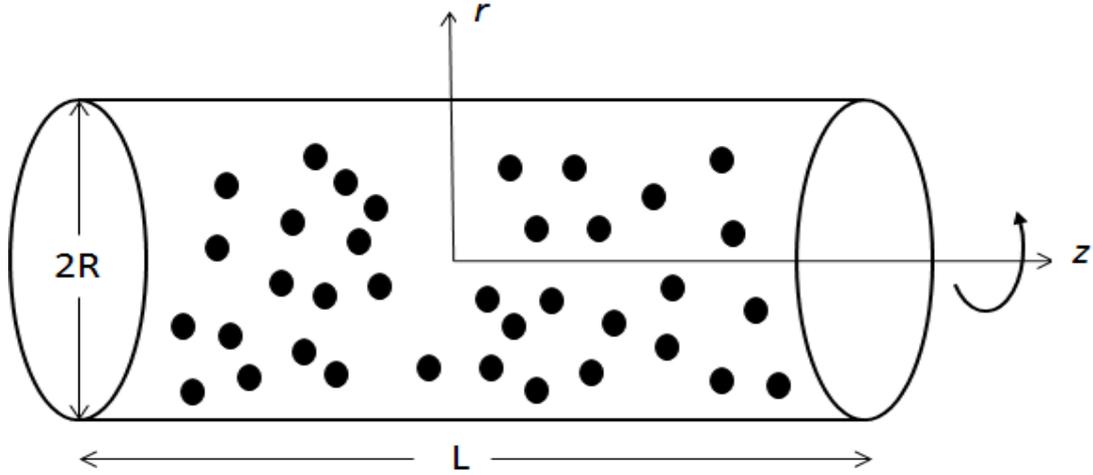

FIG. 3: Schematic representation of a rotating suspension in a horizontal cylinder.

In the Stokes regime, fluid flow field $\mathbf{u(r)}$ due to the presence of a point force $\mathbf{F}$ at $r_0$ is represented by the equation,

$$\mathbf{u(r)} = \mathscr{G}(\mathbf{r}, \mathbf{r}_0)\mathbf{F} \qquad (2.1)$$

In mathematical terms, the second order tensor $\mathscr{G}(\mathbf{r}, r_0)$ is the Green's function for the Stokes equation. The expressions for Green's function (mobility tensor) for a single particle placed along the axis of a rotating cylinder was elucidated by Liron and Shahar[26]. Lee and Ladd[24] extended the mobility tensor to N-Stokeslets in a rotating cylinder, proposing order N algorithms for both the source field and the cancelling field. On evaluation of the Green's function, the particle velocity can be represented as the sum of the contributions from the hydrodynamic interactions and the Stokes velocity as shown by the equation,

$$\mathbf{u}_j = \frac{\mathbf{F}_j}{\xi} + \sum_{i \neq j}^{N} \mathscr{G}(\mathbf{r}_j, \mathbf{r}_i)\mathbf{F}_i \qquad (2.2)$$

Neglecting the Stresslets and other higher order contributions to the force density on the particle surface eases the computational effort and can simulate approximately $10^4$ particles in most of the cases. The overall velocity field $u(r)$ can be divided into a source field $v(r)$ and a cancelling field $\omega(r)$ which can be independently calculated and the cancelling field also satisfies the equation $\omega(R) = -v(R)$ at every point on the surface of the cylinder. The source field consists of nine components $v_\beta^\alpha$ corresponding to the three directions of the Stokeslet $\alpha$ and three directions of the velocity $\beta$, here $\alpha$ and $\beta$ are components *(r, θ or z)* in cylindrical coordinates. The derivation and expressions for these components are detailed in Lee and Ladd[24]. The issues of numerical convergence such as the truncation of the infinite sum, the number of Fourier modes requisite for a specified accuracy, etc. are also addressed by Lee and Ladd[24].

Finally, from the knowledge of the particle velocities a fourth order Runga-Kutta method is used to solve the differential equation $\dot{r}_j = u_j$ so that the new positions of the particles are determined. This process is looped over until a steady state configuration is reached. All the simulations irrespective of the value of L/R have the axially periodic boundary condition (PBC) imposed in the present work.

## III. RESULTS AND DISCUSSION

Our simulations were inspired by the experimental results of Kalyankar et al.[25]. Therefore, the parameters necessary for the simulations were maintained similar to the conditions of their experiments. In our simulations the particle radius '$a$' was 100 μm and the radius of the cylinder ($R$) was $100a$. The length of the cylinder ($L$) for radial and axial segregation studies were $0.2R$ and $5R$ respectively. The fluid viscosity (μ) was 55cp (except for DB and CL cases which was 80 cp), and its density ($\rho_f$) was 1.16 g/cc. The particle density ($\rho_p$) in all but one simulation was 0.15 g/cc. Simulations were performed for various rotational frequencies (Ω). The flow Reynolds number ($Re_f = \rho\Omega R^2/\mu$) varied between 0.18 (for GB case) to 41 (for CL case).

Kalyankar *et al.* observed as many as nine independent steady states which can be distinguished by the exhibition of various radial and axial patterns. It is to be noted that in their experiments for recording observations in the radial plane a shorter cell with a diameter 1.97cm but with a length of 2.25cm was used corresponding to the average length of a single 'band' observed in the studies of axial pattern in much longer cylinders . Starting from the Granular Bed (GB) phase to the Centrifugal limit (CL) phase both the settling and buoyant systems contain a large array of identical properties and phases. Our simulations could reproduce most of the phases observed in the experimental results and to understand the dynamics of axial band formation. In the discussion below, radial and axial segregation patterns obtained from simulation results are presented exclusively.

### A) Radial Patterns

Several patterns distinct in the radial plane occur for different rotational rates of the cylinder. Simulations performed could reproduce most of the patterns that form in the radial plane for a buoyant suspension. To observe radial patterns, a cylinder with R= 1cm and L= 0.2R (to nullify axial density variations) is considered with around 2300 particles, the fluid viscosity is maintained at 55 cP at all times but for DB and CL phases the viscosity was 80cP. The density of the fluid and particles were taken to be 1.16 g/cc and 0.15 g/cc respectively. The discussion below details the steady state dynamics of the particles for a wide range of rotation rates which fall into the three regimes shown in Fig. 2.

a) Granular Bed (GB)

At very low rotational frequencies of the cylinder, the particles are stacked near the upper section of the cylinder forming a bed as shown in Fig. 4(a). This bed formation is because the dominant force in this phase is the buoyancy. As the cylinder rotates, particles from the bed are carried down along the rotating wall in a thin layer which has a thickness equivalent to the diameter of the suspended particles. At a certain downward location the vertical component of the viscous drag is overcome by the upward buoyancy force and the particles rise back into the bed with a much higher rising velocity. As the particles are dragged upwards by the buoyant force they set a clockwise current near the bed. It is evident from Fig. 4(a) that GB for the floating system is similar to that of the settling system (Fig. 4(a) in Lee and Ladd[24]) which is because the effective direction of bed formation is reversed.

b) Fingering Flow I (F1)

As the rate of rotation of the cylinder is slightly increased one can observe the elongation of the tail of GB with not much change in the properties of the bed. There is a decrease in the concentration of particles in the bed since more particles get dragged along the wall of the cylinder. The particles which are dragged down along the wall begin to move towards the bed once they reach near the bottom of the cylinder. The buoyant force which continues to increase as the particles descend becomes capable enough to lift it from the wall at the critical angular displacement. The rising particle drags its neighbouring particles as it cuts through the fluid. The region of the circulating counter currents which started to grow in the GB phase develop further in this phase owing to the increase in the rotational frequency as shown by the velocity vector plot in Fig 4 (b). The F1 phase too is identical to that of the settling suspension.

c) Fingering Flow II (F2)

F2 phase appears at slight higher velocities from F1, in this phase the particle distribution can be divided into three zones based on the circulation currents in the cylinder. It is evident from the particle distribution in Fig 4 (c) that the flow is clockwise to the left of the rotating axis as the particles that get detached from the wall tend to float back to the top. As there is no presence of particles to the right of the rotating axis, the fluid follows the course of the cylinder wall and imparts a counter-clockwise flow. These opposite currents in the left and right sections of the cylinder distort the bed further by creating another clockwise circulation near the top as shown in Fig 4 (c).

d) Low-rotation-rate-transition (LT)

This phase is a result of continuous evolution from the F2 phase as the rotational velocity is increased. A small increase in the rotational speed is enough to destroy the secondary flows associated with the fingering flow phases. The particle bed at the upper section of the cylinder which is prominent in the first three phases is completely destroyed. The circulation shown for F2 phase near the particle bed in Fig. 4 (c) becomes fully blown spreading the particles as depicted in Fig. 4 (d).

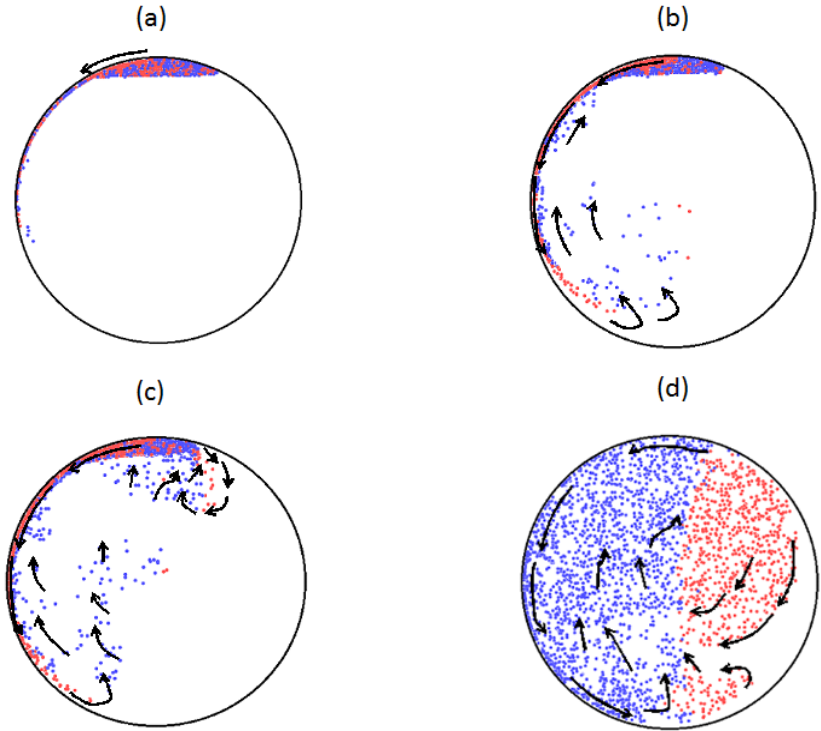

Fig 4: Velocity vectors and particle distribution at steady state for buoyancy dominated regime. Red particles move downwards whereas the blue particles move against gravity. a) $\Omega=0.09$ rad/s, particles form a bed at the top (GB) b) $\Omega=0.12$ rad/s, particles which are dragged along the cylinder tend to rise up after losing the wall influence (F1) c) $\Omega=0.15$ rad/s, more particles are dispersed into the cylinder (F1/F2) d) $\Omega=0.3$ rad/s, particles transit into the right half of the cylinder due to increased rotating velocity (LT).

Instantaneous particle positions at different times are shown in Fig. 5 to illustrate the evolution of steady state non-equilibrium patterns in the drag dominated phases from initially homogeneous distribution of particles. It is apparent that in the radial direction, the system does not take much time to attain steady state for a corresponding rotational velocity. Evolution to the GB phase illustrates the gradual movement of particles towards the top due to the dominance of drag. In the evolution to F1 phase, it can be observed that the rotation of the cylinder provides the necessary impetus for the particles to be slightly dispersed away from the cylinder. GB, F1 and F2 phases do not require too many rotations of the cylinder to reach steady state since the rotational velocity is very low. It is also obvious from the figure that the LT phase evolves to steady state configuration in much faster owing to the increased rotational velocity.

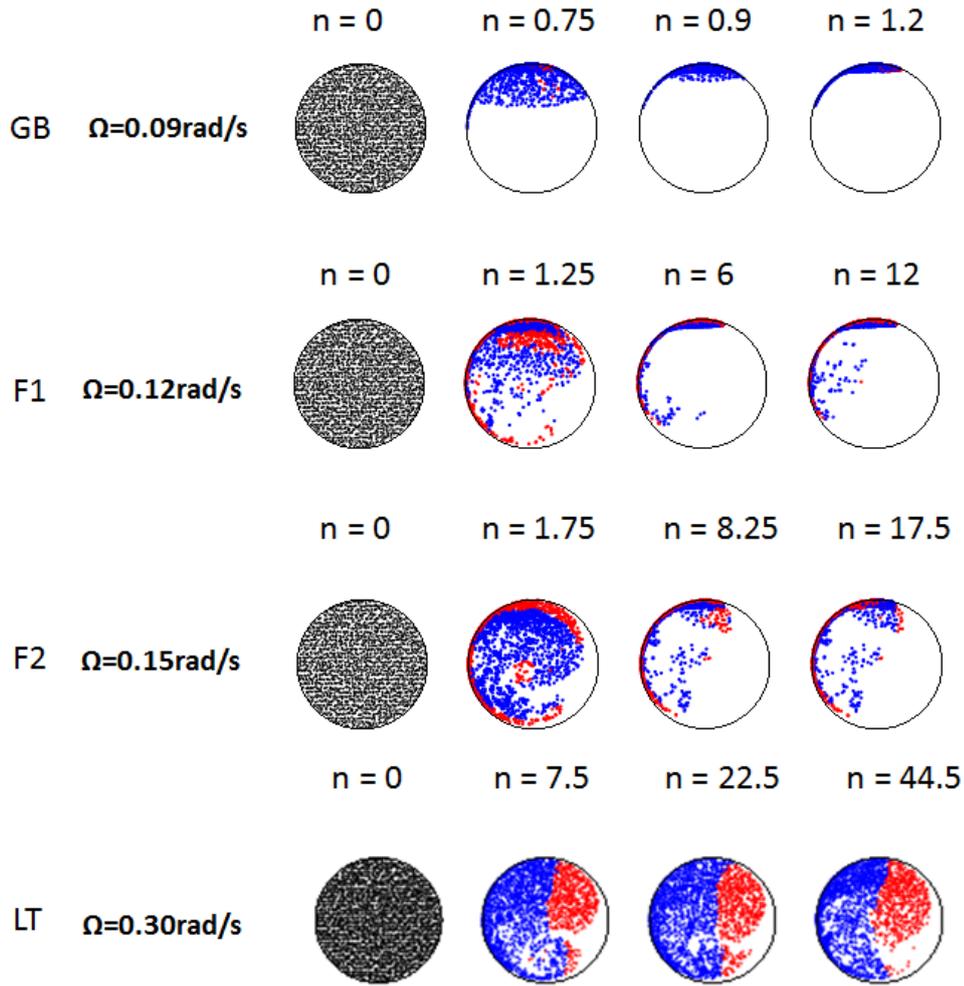

Fig 5: Phase evolution for drag dominated regime. Here 'n' represents the number of rotations of the cylinder, red particles move down and blue particles move up. Black indicates initial configuration.

e) Homogeneous Region (HR)

From Fig 6 (a) we can see that there are absolutely no secondary flows associated to the rotational flow of the suspension. This leads to almost uniform distribution of particles along the radial plane. The bed formed due to particle aggregation near the top due to buoyancy in the previously observed phases is destroyed completely. Velocity vectors and particle distribution indicate that the flow is equivalent to rigid body rotation as majority of particle trajectories have a fixed axis of rotation. The occurrence of the HR phase coincides with the balance between the buoyancy and centrifugal forces since the phases before this show the dominating nature of buoyant force while the phases which occur after HR indicate that centrifugal forces have stronger influence.

f) Discontinuous Banding (DB)

Fig 6 (b) shows the radial pattern observed at 15 rad/s. Since the particles are less dense than the suspending fluid centrifugal force draws the particles radially towards the centre of the cylinder. As these particles are drawn towards the centre of the cylinder, they tend to form a core around the rotational axis with the remaining particles forming a cloud around the central core.

g) Centrifugal Limit (CL)

In the case of settling suspension the particles swirl out to the wall of the cylinder, whereas the particles in this system form a cluster around the rotating axis of the cylinder. Our simulations reproduced the fact that the fraction of particles which constitute the central core is ~0.6 of the total number. Direction of motion of the particles indicated by arrows show similarity in behaviour to the DB phase as the qualitative behaviour of particles remains the same in both the phases.

Fig. 7 gives the evolution of steady state non-equilibrium patterns in the centrifugal force dominated phases. It is clearly observed that the qualitative behaviour of the particles in DB and CL phases is mostly similar. Change in the direction of the particle motion (as to where they settle at steady state) owing to the increase in the rotation rates is evident. Moreover, we can also understand that increase in the rotation rates in turn tends to increase the centrifugal forces as particles show more affinity to form clusters at the centre of the cylinder. The increase in the number of rotations of the cylinder to attain steady state is because it takes a while to form a dense core of particles.

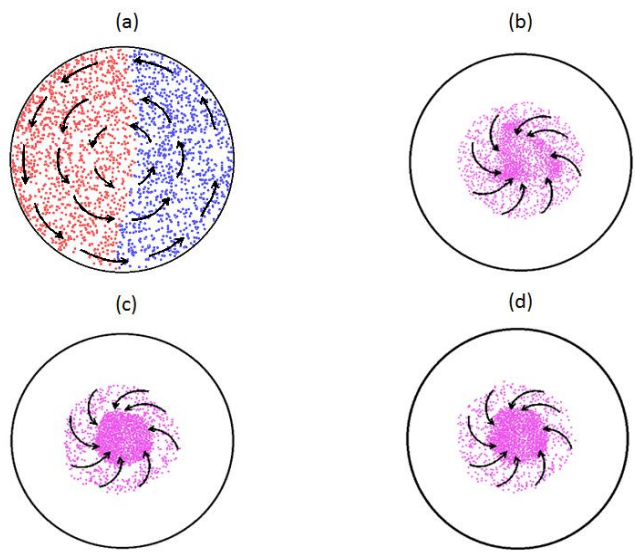

Fig 6: Velocity vectors and particle distribution at steady state for the centrifugal force dominating regime. Red particles move downwards whereas the blue particles move against gravity. Magenta indicates radially inward motion. a) Ω=0.75 rad/s, particles are dispersed throughout the cylinder rotating along with it (HR)  b) Ω=15 rad/s, increased centrifugal force propels particles towards the rotating axis (DB) c) Ω=30 rad/s, (CL) particles congregate around the centre of the cylinder  d) Ω=35 rad/s, (CL) further increased velocity produces no change in the qualitative behaviour of the particles.

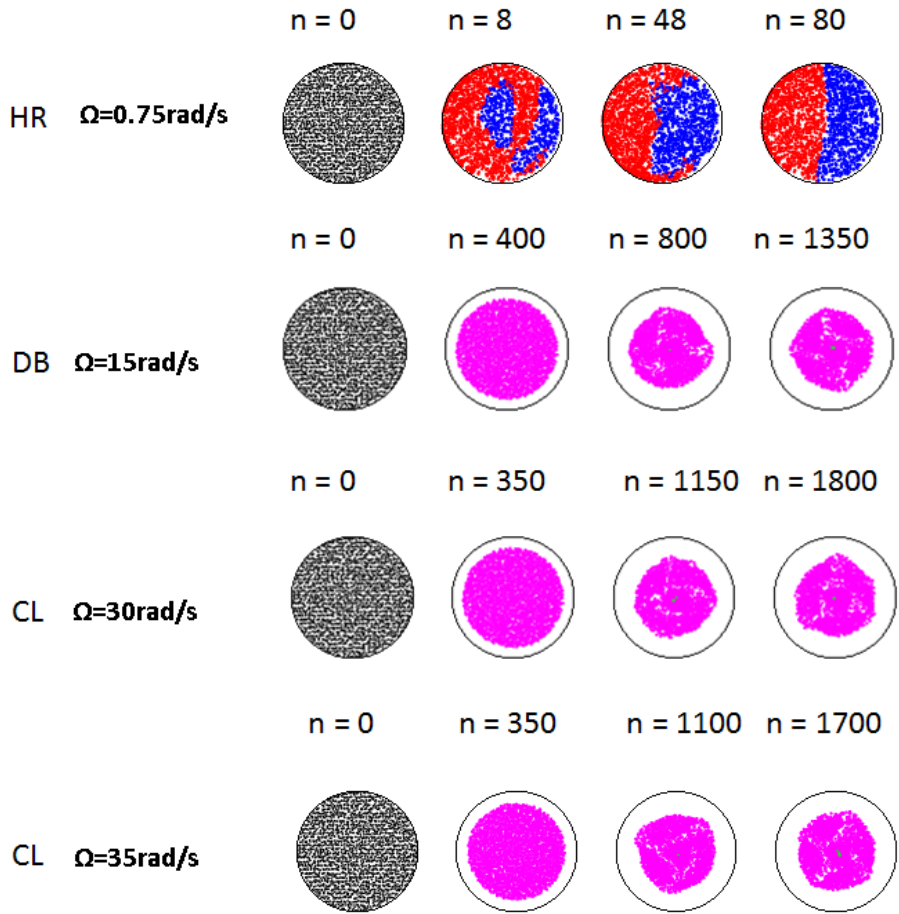

Fig 7: Phase evolution for centrifugal force dominated regime. Black colour is used for the initial configuration. Here 'n' represents the number of rotations of the cylinder, red particles move down & blue particles move up. Magenta indicates inward motion towards the centre of the cylinder.

Fig. 8 shows the radial-concentration profile at different frequencies along the positive y-axis. In the GB phase the buoyancy force is much larger compared to the viscous drag and centrifugal forces. This causes all the particles to rise up and accumulate near the top. For the HR phase the profile is almost flat suggesting nearly homogeneous distribution of particles throughout the cylinder. In this regime the magnitude of different forces is comparable. The DB and CL phase profiles indicate the build-up of particle beds along the rotating axis of the cylinder.

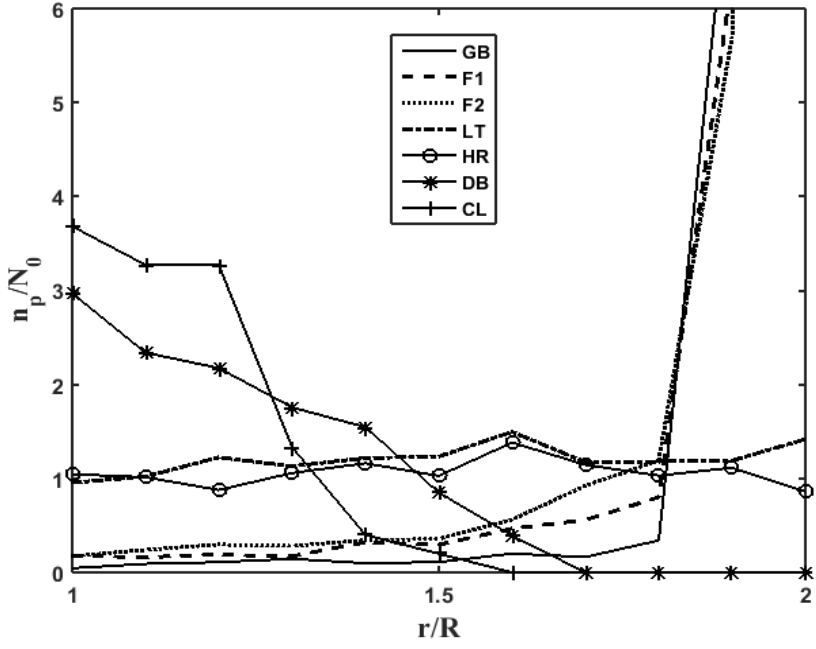

FIG. 8: Equilibrium concentration profiles in the radial direction for different frequencies of rotation. Here $n_p$ and $N_0$ are the number of particles after reaching steady state and at t = 0 respectively.

B) Order Parameters

The particle distribution and the direction of particle motion from Fig. 4 suggest that the order parameter can be defined as $\langle \dot{\theta} \rangle = \langle \sum_{i=1}^{N} \dot{\theta}_i \rangle / N$ based on the time-averaged angular velocity. It could distinguish between the drag dominating segregated phases where the particle returns to the bed soon after getting detached from it and the homogeneously distributed. The order parameter given by the expression $Q = \dfrac{\langle \dot{\theta} \rangle}{\Omega}$ was defined by Lee and Ladd[22] for the sedimenting particle system. Since the experimental conditions for both the settling and the floating particle systems are identical, the postulation of Lee and Ladd[22] could be used for the present work. The significant quantities which characterise Q are the velocity of the fluid, rising velocity of the particle ($u_f = m_B g \xi^{-1}$) as defined in the single particle dynamics; $l$ being the characteristic length. Therefore, flow under the experimental conditions can be categorized by the dimensional ratios $\dfrac{u_f}{\Omega l}$ and $\dfrac{u_c}{\Omega l}$. As discussed earlier all phases prior to the occurrence of the homogeneous region fall under the buoyancy/gravity dominated regime and the phases which follow after HR fall under the influence of centrifugal force due to the rotational velocity of the cylinder. Hence the dimensionless ratio $\frac{u_f}{\Omega l}$ is used in the determination of the order parameter for the buoyancy dominated regime. Fig. 9 shows the variation of the order parameter Q with the rotational frequency of the drum. Fig. 9(a) shows that the size of the cylinder has no influence on the transition frequency. However it can be seen that the mean particle concentration affects the transition frequency, as it affects the mean rising velocity of particles in the buoyancy dominated regime. The numerical value of Q lies between 0 and 1 each representing completely segregated and fully dispersed phases. In Lee and Ladd[22] it is suggested that for a settling system, the mean interparticle separation $d = n_0^{-1/3}$ is the characteristic length, where $n_0$ is the average particle concentration. Fig. 9(b) affirms the claim that even for the floating system the mean interparticle separation is the characteristic length. Choice of mean interparticle distance for the characteristic length is supported mathematically by a mass balance over the low rotational frequency phases. The up-flux of the particles due to buoyancy scales as $\dot{M} \sim u_f n R L$ (here $n$ is the number of particles in the monolayer) whereas the down-flux of a monolayer of particles getting dragged down by the rotation of the cylinder scales as $\dot{M} \sim \Omega a n R L$ (monolayer of particles have a thickness proportional to their radius '$a$'). A balance over these fluxes shows that the resulting dimensionless number is $\Omega a / u_f$.[22]

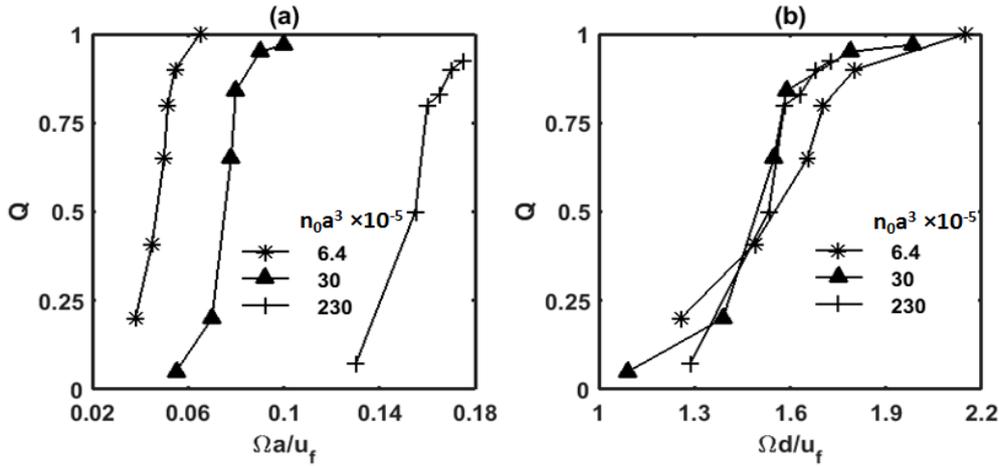

FIG. 9: Order parameter (Q) for various particle concentrations, asterisks is plotted versus the two dimensionless numbers in (a) $\Omega a / u_f$ and in (b) $\Omega d / u_f$, where $d = n_0^{-1/3}$

C) Axial Segregation

To observe axial segregation, simulations were performed with $L=5R$ and with a total of 12435 particles. All the studies in this section are performed with a random initial configuration of particles as shown in Fig. 10(a). Several distinct phases such as HR, LD, SB and CL observed experimentally were reproduced through the simulations. HR shows complete mixing whereas LD and SB exhibit axial particle bands. Fluctuations of particle density in the radial plane induce additional movements of particles in the axial direction. These additional movements in the axial direction serve as perturbation to the axial particle density. As steady state is reached the perturbations grow into large axial density variations thereby leading to the formation of axial banding patterns. The observations are detailed in the description of LD and SB phases.

a) Homogeneous Region (HR)

From Fig. 10(b) it is evident that the particle distribution in the cylinder shows uniformity. The particle clusters which remain near the wall which appear for GB, F1 & F2 phases is completely destroyed as the particles mix uniformly. The influence of drag seen in LD is absent in this phase as the particle motion becomes identical to rigid body rotation. The particles can be seen to be moving both in the positive and negative z-direction, signifying particle mixing leading to a homogeneous distribution.

b) Local-structure Dropout (LD)

Unlike the previous phases the LD does not show any form of resemblance to the settling system. The settling system which is characterized by the redistribution of particles spreads from one location along the axial direction but for the buoyant system there is spatiotemporal chaos. The phase is characterized by exchange of particles between the bands and oscillation in bands with time is observed as a consequence. This oscillation of bands and their non-uniform structure is well illustrated in Fig.11; particle concentration for different number of rotations indicates the drifting mechanism experienced by the particles. In this phase, the centrifugal forces are not as weak as in the case of phases prior to the HR phase. As the particles are less dense than the fluid, these particles are pulled in towards the centre of the cylinder. The interplay between a relatively strong centrifugal force and the buoyant force cause these particles to apparently spiral inwards and outwards causing exchange of particles and oscillation of bands. This form of oscillation in bands is similar with the phenomenon reported by Breu *et al.*[17]. The axial patterns shown in the associated Fig. 10(c) and Fig. 10(d) are obtained after 750 rotations of the cylinder at 2.5 rad/s.

c) Centrifugal Limit (CL)

On the occurrence of the Centrifugal Limit phase, most of the particles concentrate around the axis of rotation of the cylinder. Along the axial direction Fig. 12 shows the central particle core which has a fraction (~ 0.6) of the total particles, while the rest of the particles form a cloud surrounding the central core. However, the simulations were unable to reproduce bands in the DB phase and during the transition from DB to CL phase which were reported experimentally.

These observations could lead us to a conclusion that hydrodynamic interactions involved in the dilute system influence particles closer to each other to develop a buckling instability which leads into clustering of particles. It can now be firmly implicated that the axial differences in particle concentration are catalysed by these particle clusters which become denser as more particles get drawn into the already formed clusters. The curved cross-section of the rotating vessel also enhances the growth of this instability which results in axial banding. This is reinstated by the fact that the fluid down-current due to the rotation of the drum enhances the rate at which particles rise to the top of the cylinder. Nevertheless, occurrence of the HR phase in between phases where axial banding was evident implies that the balance among forces acting on the suspension and the cross-section of the cylinder are pivotal for the instability to develop. Any imbalance among these forces leads to a buckling instability causing periodic clustering of particles in the cylinder. Thence imbalances among the centrifugal and buoyancy (gravity and drag) cause a variation in the radial concentration of particles eventually leading into particle clustering and banding along the axial direction. Both numerical and experimental results re-establish that there would be no axial concentration bands when there is homogeneity in the concentration of particles in the radial plane.

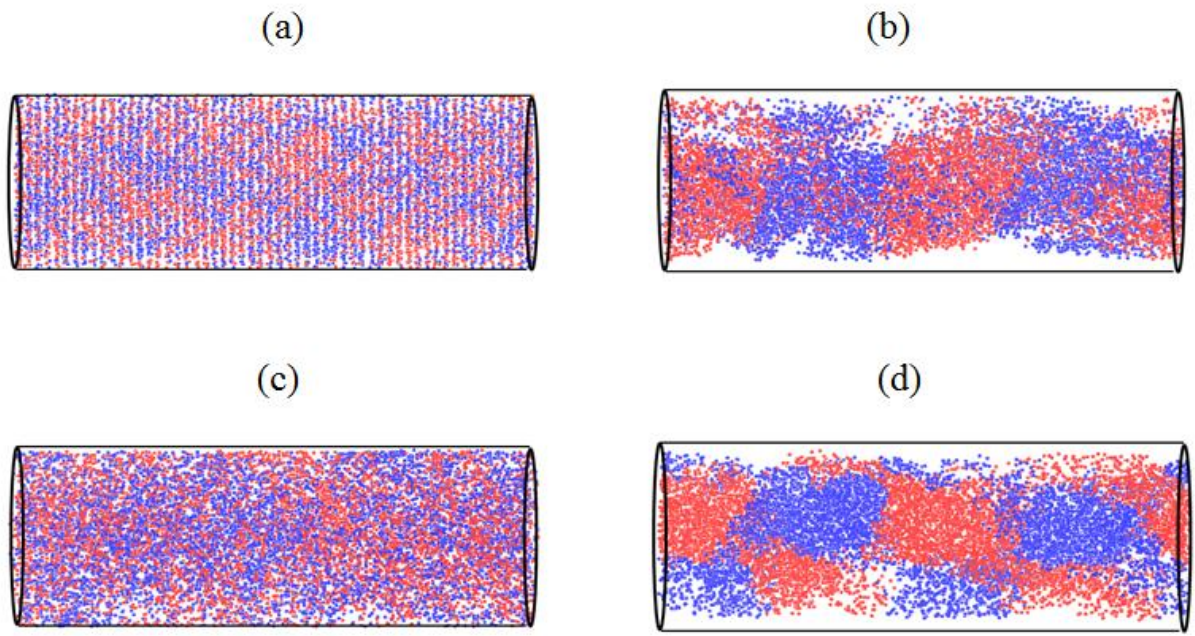

FIG 10: Axial patterns observed in the centrifugal force dominated regime. Red particles move towards the right and the blue particles move leftwards. a) random initial configuration of particles suspended in the cylinder b) HR phase, $\Omega$ = 1.25rad/s, uniform distribution of particles throughout the cylinder. c) LD, $\Omega$ = 2.5rad/s (top view, gravity is pointing into the plane of the paper): Particles converge into high-concentration regions while rising and spread out as they reach the top. d) LD, $\Omega$ = 2.5 rad/s (front view, gravity is pointing downward).

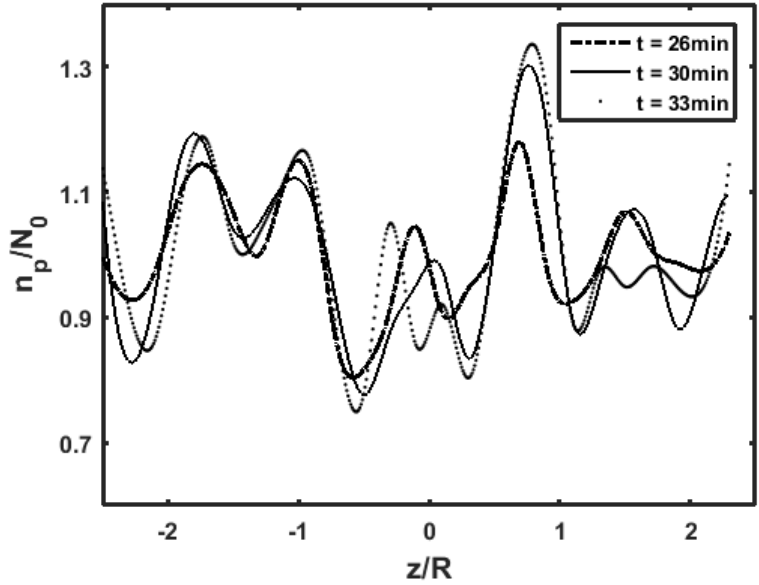

FIG. 11: Number density at three different instants indicating exchange of particles along the axial direction for particle density, $\rho_p$ = 0.15g/cc and $\Omega$ = 2.5 rad/s. Here $n_p$ and $N_0$ are the number of particles after reaching steady state and at t = 0 respectively.

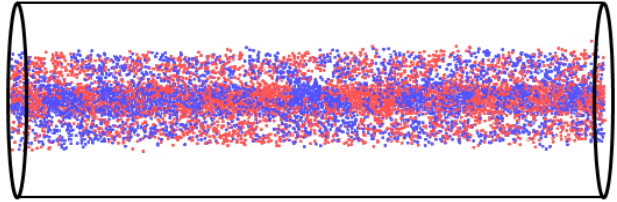

FIG 12: Central core formation in the Centrifugal Limit (CL) $\Omega$ = 30 rad/s. Particles congregate around the rotational axis of the cylinder reaching the maximum packing limit for a suspension.

d) Stable Bands (SB)

We have carried out one simulation where the density of particle was increased to 0.19 g/cc and the rotational frequency was 2.1 rad/s. Fig. 14 shows the plot of particle number density along the axial direction. Unlike the travelling bands in the LD phase (Fig. 11) the stripes of particles in SB phase remain stationary as steady state is reached. This indicates that in the Stable band (SB) phase there may not be any exchange of particles between consecutive bands. This reinforces the statement made earlier that the traveling of bands is observed only for certain imbalance between buoyancy and centrifugal forces. It is also evident that the number of bands is increased to three from the LD phase ($\rho_p$ = 0.15g/cc, $\Omega$ = 2.5 rad/s) which contains only two bands. We are unable to provide any explanation for the difference in the number of bands between the LD and SB phase.

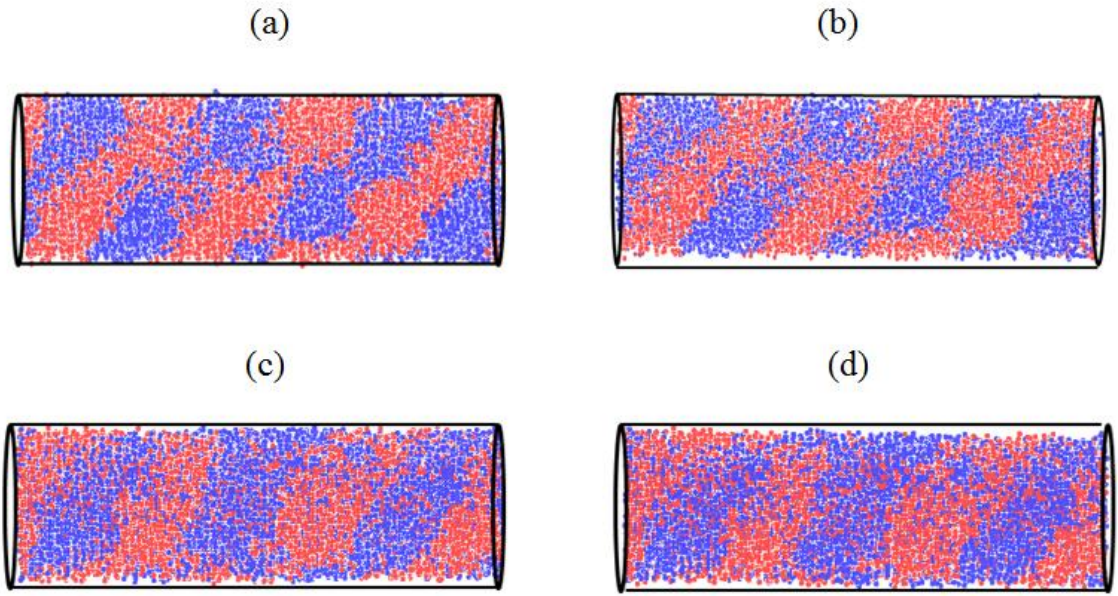

FIG 13: Axial segregation ($\rho_p$ = 0.19g/cc) in the stable band phase. (a) & (b) correspond to front view and top view at 800 rotations of the cylinder correspondingly (c) & (d) are at 900 rotations at $\Omega$ = 2.25 rad/s. Red particles move towards the right and the blue particles move leftwards.

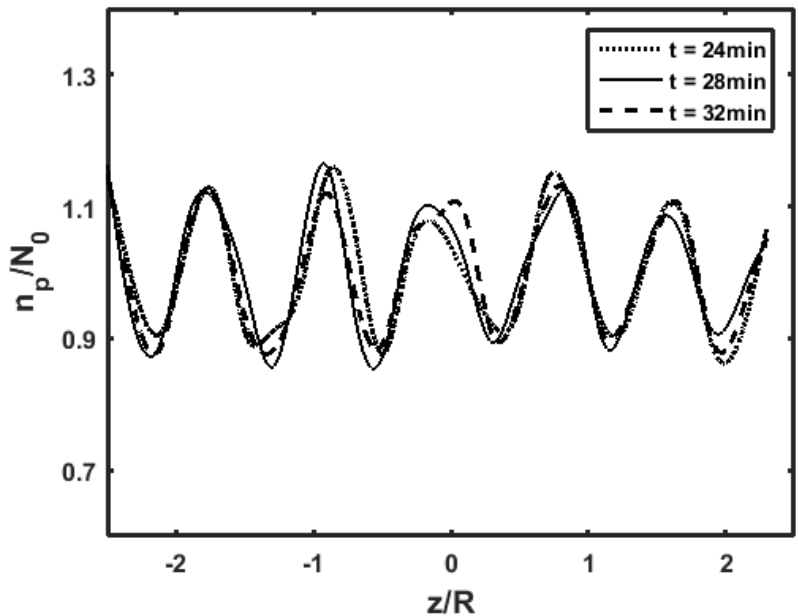

FIG 14: Number density of particles for $\rho_p$ = 0.19 g/cc at $\Omega$ = 2.1 rad/s. There is almost no exchange of particles along the rotating axis over time indicating stable band patterns. Here $n_p$ and $N_0$ are the number of particles after reaching steady state and at t = 0 respectively.

## IV. COMPARISON WITH EXPERIMENT

As reported earlier nine independent non-equilibrium states were observed in the experiments of Kalyankar et al.[25] for positively buoyant suspensions. Our simulation studies could reproduce the characteristic behaviour of almost all the phases. For the three regimes classified earlier, the radial patterns were reproduced quite convincingly. The axial banding patterns observed in the SB and LD phase have both qualitative and quantitative agreement with the experimental results. However, the axial bands for the high frequency DB phase and also in the low frequency regime for the GB/F1 phases, the simulations do not capture completely the experimentally reported behaviour. These discrepancies with the experimental observations evidently occur when the concentration of the particle is relatively high. In the simulations the particles are approximated as Stokeslets and the contribution from higher order multipoles arising from the finite size of the particle are neglected. Reasonable agreement of radial patterns was observed for GB and CL phases where the particle concentration in certain regions goes up. In the granular bed regime, though the particle concentration is very high, there is not much movement of the particles, and the hydrodynamic drag is weaker compared to the buoyancy force. On the other hand, in case of centrifugal limit, though the particle velocities are higher the radially symmetric distribution may ensure that the results from point particle simulations would be qualitatively similar to the case of finite sized particles at higher volume fractions.

Recent studies performed by Refs. 28-30 incorporated higher order multipolar solutions to analyse the case of many-body dynamics inside a cylinder by applying proper no-slip boundary conditions on the particle surface. Unlike the case of point particles, inclusion of higher order multipoles takes into account the effects of the finite size of the particle. Addition of the effects of the particle size would probably capture the dynamics in the low and high frequency regimes where the phases exhibit high concentration as these interactions are primarily due to mutual hydrodynamic influence or direct contact in collision. These short-range lubrication forces are expected to cause enhanced diffusivity as indicated by Zurita-Gotor et al.[31]. The enhanced diffusivity is a prime factor for the observed difference in the value of the dimensionless frequency $\Omega^*(=\Omega d/u_f)$ which ranges between $\{1.1 - 1.5\}$ for experiments and $\{1.35 - 1.75\}$ for simulations. The deviations of the simulation results from experimental observations and the axial patterns in the DB phase arise due to the exclusion of Stresslets and lubrication forces. Another reason for the deviation from experimental results could be due to the periodic boundary conditions in the simulations which is different from the experimental boundary conditions of no-slip rotating end-walls. It can be expected that the no slip rotating end-wall might strengthen the rigidly rotating fluid field near the wall. However, the simulation methodology is currently not equipped with providing end-wall with no slip condition.

The convective contribution to band formation might also be a factor for the band formation in DB phase but it has no effect on the low frequency phases. Inclusion of convective effects may capture the dynamics of particles in the DB phase and the study might as well be extended to understand the patterns reported by Lipson[15] and Breu et al.[16]

## V. CONCLUSION

The present work is an application of the method proposed by Lee and Ladd[22] to understand the formation of axial bands for a buoyant particle system experimented by Kalyankar et al.[25]. Most of the distinct patterns which appeared in both radial and axial directions obtained by Kalyankar et al.[25] were reproduced by this simulation technique which includes only the far-field hydrodynamic interactions. The oscillation of bands with time in the LD phase was also observed. The order parameter Q and the characteristic length $l$ are comparable with the settling system of Lee and Ladd[24] in the buoyancy dominated regime. Axial banding observed by Kalyankar et al.[25] in the DB phase could not be reproduced in the simulations.

According to Lee and Ladd[21], buoyant system would not be able to produce axial patterns as there would be repulsive interactions among floating particles unlike settling particles which always have attractive interactions. However, this theory seriously underestimated the strength of cancelling field with a flat wall approximation. Although the source field interaction should be indeed attractive or repulsive for heavier or lighter particles than the medium, the cancelling field exactly nullifies the source field interaction between ring stokeslets along the axial direction. Both experiments and simulations could recover banding patterns in the axial direction for a buoyant particle system. This can be ascribed to the inhomogeneous distribution of particles in the radial plane due to imbalance in the forces acting on the particle as explained for settling particles by Lee and Ladd[24].

Though there is a difference in the direction of centrifugal force for the settling and floating particle suspensions, it is clear that axial banding phenomenon is observed in either system. Simulation results show that to achieve complete mixing of particles there must be perfect balance of all the forces acting on them. Therefore it is apparent that the axial bands are formed only when a there is a certain imbalance in the forces acting on the particles. This imbalance in both the systems changes due to the change in the direction of the centrifugal force causing the appearance of HR phase before the SB phase in the buoyant system. If we recall from the single particle dynamics, HR phase is apparently a balance between the forces in the system. As the domination of forces shifts from buoyant to centrifugal there is formation of HR phase and hence the shift in position from the settling system.

ACKNOWLEDGEMENT

We would like to thank IIT Guwahati for providing the HPCC facility.